\documentclass[submission,copyright,creativecommons]{eptcs}
\providecommand{\event}{WORDS 2011} % Name of the event you are submitting to
\usepackage{breakurl}             % Not needed if you use pdflatex only.
\usepackage{amssymb}
\usepackage{amsmath}
\usepackage{amsfonts}
\newtheorem{theorem}{Theorem}
\newtheorem{lemma}{Lemma}

\title{Permutation complexity of the fixed points of some uniform binary morphisms}
\author{Alexander Valyuzhenich
\institute{Novosibirsk State University, Novosibirsk, Russia}
\email{graphkiper@mail.ru}
}
\def\titlerunning{Permutation complexity of the fixed points of some uniform binary morphisms}
\def\authorrunning{Alexander Valyuzhenich}
\begin{document}
\maketitle

\begin{abstract}
An infinite permutation $\delta$ is a linear order on the set $\mathbb{N}$.
We study the properties of infinite permutations generated by fixed points of some uniform binary morphisms, and find the formula for their complexity.
\end{abstract}

\section{Introduction}
The notion of an infinite permutation was introduced in \cite{flaas}, where were investigated the periodic properties and low complexity of permutations.
Similarly to the definition of subword complexity of infinite words, we can introduce the notion of the factor complexity of a permutation as the number of distinct subpermutations of a given length.
The notion of a permutation generated by an infinite non-periodic word was introduced in \cite{mak}.
In \cite{sturm} Makarov calculated the factor complexity of permutations generated by a well-known family of Sturmian words.
In \cite{widmer} Widmer calculated the factor complexity of the permutation generated by the Thue-Morse word.
In this paper we find a formula for the factor complexity of permutations generated by the fixed points of binary uniform morphisms from a some class. Since the Thue-Morse word belongs to this class, we obtain an alternative way to compute the factor complexity of the Thue-Morse permutation.

In Section 2 we introduce basic definitions to be used below.
In Section 3, we introduce the class of morphisms $Q$ for which in Section 9 we state the main theorem of this paper.
In Sections 4--8 we state some auxiliary assertions needed to prove the main theorem.
In Section 10 we give an alternative proof of the formula for the factor complexity of the permutation generated by the Thue-Morse word.

\section{Basic definitions}
Let $\Sigma$ be a finite alphabet. Everywhere below we will use only the two-letter alphabet $\Sigma=\{0,1\}$.

A right infinite word over the alphabet $\Sigma$ is a word of the form $\omega=\omega_{1}\omega_{2}\omega_{3}\ldots$, where each $\omega_{i}\in{\Sigma}$. A (finite) word $u$ is called a subword of a (finite or infinite) word $v$ if $v=s_{1}us_2$ for some words $s_1$ and $s_2$ which may be empty. The set of all finite subwords of the word $\omega$ is denoted by $F(\omega)$.
 For the word $\omega$ we define the binary real number $R_{\omega}(i)=0.\omega_{i}\omega_{i+1}\ldots=\sum_{k\geq{0}}{\omega_{i+k}2^{-(k+1)}}$.
The mapping $h:\Sigma^{\ast}\longrightarrow{\Sigma^{\ast}}$ is called a morphism if $h(xy)=h(x)h(y)$ for any words $x,y\in{\Sigma^{\ast}}$.
We say that $\omega$ is a {\em fixed point} of a morphism $\varphi$ if $\varphi(\omega)=\omega$.
Clearly, every morphism is uniquely determined by the images of symbols, which we call {\em blocks}.
A morphism is called {\em uniform} if its blocks are of the same length.

We say that a morphism $\varphi:\Sigma^{\ast}\longrightarrow{\Sigma^{\ast}}$  is {\em marked} if its blocks are of the form $\varphi(a_i)=b_ixc_i$, where $x$ is an arbitrary word, $b_i$ and $c_i$ are symbols of the alphabet $\Sigma$, and all $b_i$ (as well as all $c_i$) are distinct.
In what follows, we will consider only uniform marked morphisms with blocks of length $l$.

An {\em interpretation} of a word $u\in\Sigma^*$ under the morphism $\varphi$ is a triple $s=\langle{v,i,j}\rangle$, where $v=v_{1}\ldots{v_{k}}$ is a word over the alphabet $\Sigma$, $i$ and $j$ are nonnegative integers such that $0\leq{i}<|{\varphi(v_1)}|$ and $0\leq{j}<|{\varphi(v_k)}|$, and the word obtained from $\varphi(v)$ by erasing $i$ symbols to the left $i$ and $j$ symbols to the right is $u$.
Moreover, if $v$ is a subword of $\omega$, then $s$ is called an interpretation on $\omega$.
The word $v$ is called an {\em ancestor} of the word $u$.
In what follows we shall consider only interpretations of subwords of the word $\omega$.
We say that $(u_1,u_2)$ is a {\em synchronization point} of $u\in{F(\omega)}$ if $u=u_1u_2$ and $\forall{v_1, v_2}\in{\Sigma^{*}},\forall{s}\in{F(\omega)}$ $\exists{s_1,s_2}\in{F(\omega)}$ such that  $[v_{1}uv_{2}=\varphi(s)\Rightarrow(s=s_{1}s_{2},v_{1}u_{1}=\varphi(s_1),u_{2}v_{2}=\varphi(s_2))]$. A fixed point $\omega=\varphi(\omega)$ of the morphism $\varphi$ is called {\em circular} (\cite{frid}), if any subword $v$ of word $\omega$ of length at least $L_{\omega}$ contains at least one point of synchronization.

For uniform morphisms, this means the uniqueness of the partition of the word $v$ into blocks.

In $\cite{frid}$ it was proved that the nonperiodic fixed points of uniform binary morphisms with $w_1=0$ are circular, except for the case when $\varphi(1)=1^n$.

 An {\em occurrence} of word $u\in\Sigma^*$ in the word $\omega$ is a pair $(u,m+1)$ such that $u=\omega_{m+1}\omega_{m+2}\ldots{\omega_{m+n}}$.
It is easy to see that a word can have many different occurrences.

Since the fixed point $\omega$ is circular, the interpretations of all occurrences of the word $u$ are the same and equal to $\langle{v,i,j}\rangle$.

An occurrence $(v,p+1)$ of a word $v$ of length $k$ is called the {\em ancestor of the occurrence} $(u,m+1)$ of the word $u$ if $m=pl+i$ and $0\leq{i}<l$.

It is easy to see that for $|u|\geq{L_{\omega}}$ word $u$  has exactly one ancestor, and any occurrence $(u,m+1)$ of the word $u$ also has exactly one ancestor.

Let $|u|\geq{L_{\omega}}$.
A sequence $u_{0},u_1,\ldots{,u_m}$ of subwords of word $\omega$ is called a {\em chain of ancestors} of the word $u$, if $u_{i+1}$ is the ancestor of $u_i$ for any $0\leq{i}\leq{m-1}$ and $u_0=u$.

The chain of ancestors of word $u$ will be denoted as $u\rightarrow{u_1}\rightarrow{\ldots}\rightarrow{u_m}$.

We say that $u$ is a {\em descendant} of $v$ if $v$ belongs to a chain of ancestors of $u$.

 Now we introduce the main object of this paper.

Let $\omega$ be a right infinite nonperiodic word over the alphabet $\Sigma$.

We define the {\em infinite permutation} generated by the word $\omega$ as the ordered triple $\delta=\langle\mathbb{N},{<}_{\delta},<\rangle$, where ${<}_{\delta}$ and $<$ are linear orders on $\mathbb{N}$.

The order ${<}_{\delta}$ is defined as follows: $i{<}_{\delta}j$  if and only if $R_{\omega}(i)<R_{\omega}(j)$, and $<$ is the natural order on $\mathbb{N}$.

Since $\omega$ is a non-periodic word, all $R_{\omega}(i)$ are distinct, and the definition above is correct.

We define a function $\gamma:R^{2}\rightarrow\{<,>\}$, which for two different real numbers reveals their relation.

We say that a permutation $\pi=\pi_{1}\ldots{\pi_n}\in{S_n}$ is a {\em subpermutation} of length $n$ of an infinite permutation $\delta$ if $\gamma(\pi_s,\pi_t)=\gamma(R(i+s),R(i+t))$ for $1\leq{s}<{t}\leq{n}$ and for a fixed positive integer $i$.

We define the set $Perm(n)=\{\pi(i,n+i-1)|i\geq{1}\}$, where $\pi(i,n+i-1)=\pi_{i}\ldots{\pi_{n+i-1}}$ is subpermutation induced by the sequence $R(i),\ldots,R(n+i-1)$  (in the sense that $\gamma(\pi_{i+s_1},\pi_{i+s_2})=\gamma(R(i+s_1),R(i+s_2))$ for $0\leq{s_{1}}<s_{2}\leq{n-1}$).

Now we define the {\em permutation complexity} of the word $\omega$ (or equivalently, the factor complexity of the permutation $\delta_{\omega}$) as $\lambda(n)=|Perm(n)|$.
We say that an occurrence $(u,m+1)$ of the word $u$ {\em generates} a permutation $\pi$ if $\pi$ is induced by a sequence of numbers $R(m+1)\ldots{R(m+n)}$.
A subword $u$ of the word $\omega$  {\em generates} the permutation $\pi$ if there is an occurrence  $(u,m+1)$ of this word which generates $\pi$.

\section{Morphisms considered}
We say that a uniform marked binary morphism $\varphi$ with blocks of length $l$ belongs to the class $Q$ if one of the following conditions is fulfilled: either $\varphi(0)=X=01^{n}0x1,\varphi(1)=Y=10^{m}1y0$, where ${n,m}\in\mathbb{N}$, both the word $1^{n}$ and the word $0^{m}$ is included in both blocks morphism exactly once, and the word $X$ ($Y$) does not end by $1^{n-1}$($0^{m-1}$); or $\varphi(0)=01^{n},\varphi(1)=10^{n}$, where $n=l-1$.

It is easy to see that the fixed point $\omega=\lim\limits_{n \to\infty}{\varphi^{n}(0)}$ of any morphism $\varphi$ which belongs to the class $Q$ is circular because $\varphi(1)\neq{1^n}$ for any $n$.

Everywhere below the word $\varphi(0)$ will be called the block of the first type, and $\varphi(1)$ is called the block of the second type.

\textbf{Example.} The morphism $\varphi(0)=011101,\varphi(1)=100010$ belongs to $Q$, whereas the morphism $\varphi(0)=01011,\varphi(1)=10000$ does not belong to $Q$.

Consider a fixed point $\omega=\varphi(\omega)$ of a morphism $\varphi\in{Q}$. Then $\omega$ is divided into blocks, which are the images of its symbols. Such a partition is called {\em correct}.

\begin{lemma}
Let $\omega$ be a fixed point of the morphism $\varphi$, where $\varphi\in{Q}$. Then the following statements are true:

\begin{enumerate}
  \item Let $\omega_{i}=\omega_{j}=0$ and $i\equiv 1 \bmod l$, $j\not\equiv 1 \bmod l$. Then $R_{\omega}(i)>R_{\omega}(j)$.
  \item Let $\omega_{i}=\omega_{j}=1$ and $i\equiv 1 \bmod l$, $j\not\equiv 1 \bmod l$. Then $R_{\omega}(i)<R_{\omega}(j)$.
\end{enumerate}

\end{lemma}

\begin{lemma}
 Let $\omega$  be a fixed point of the morphism $\varphi$, where $\varphi\in{Q}$. Let $\omega_{i}=\omega_{j}$, where $i\equiv{i'}$$(mod\ l)$,$j\equiv{j'}(mod\ l)$ and $0\leq{i',j'}\leq{l-1}$, where $i'$ and $j'$ are fixed. If $i'\neq{j'}$, or if $\omega_i$ and $\omega_j$ lie in blocks of different types in the correct partition $\omega$ into blocks, then the relation ${\gamma(R_{\omega}(i),R_{\omega}(j))}$ is uniquely defined by $i'$, $j'$ and the types of respective blocks.

\end{lemma}

\begin{lemma}
Let $\omega_{i}=\omega_j$ and $R_{\omega}(i)<R_{\omega}(j)$. Then inequality $R((i-1)l+r)<R((j-1)l+r)$ holds for all $1\leq{r}\leq{l}$.

\end{lemma}

\section{Equivalence of permutations}
In this section we introduce the concept of equivalent permutations. Let $z=z_{1}z_{2}\ldots z_{k}$ be a permutation belonging to $S_k$.

An {\em element} of the permutation $z$ is the number $z_i$, where $1\leq{i}\leq{k}$.

We will say that two permutations $x=x_{1}x_{2}\ldots x_{k}$ and $y=y_{1}y_{2}\ldots y_{k}$ are {\em equivalent}  if they differ only in relations of extreme elements, i.e $\gamma(x_{1},x_{k})\neq{\gamma(y_{1},y_{k})}$, but $\gamma(x_{i},x_{j})={\gamma(y_{i},y_{j})}$ for all other $i,j$. We will denote this equivalence by $x\sim{y}$.

\begin{lemma}
Let $x$ be a finite permutation and $x=x_{1}\ldots,x_{k}$. Then the permutation $y$ such that $x\sim{y}$ exists if and only if $|x_{1}-x_{k}|=1$.
\end{lemma}

\section{Bad, narrow and wide words}
 For an arbitrary subword $v$ of the word $\omega$ we define the sets $M_{v}$ and $N_{v}$, where $N_{v}$ is the set of all pairs of equivalent permutations, and $M_v$ is the remaining set of permutations generated by $v$.

A word $u$ will be called {\em bad} if the set $N_u$ is not empty, i.e, if $u$ generates at least one pair of equivalent permutations.

Let $u$ be a subword of word $\omega$ of length $|u|=n$. The number of permutations generated by $u$ is denoted by $f(u)$.

\begin{lemma}
Let $u$ be a word of length $|u|=n\geq{L_{\omega}}$, and $u'$ be the ancestor of $u$. Then the following statements are true:

\begin{enumerate}
  \item $f(u)\leq{f(u')}$,
  \item If  $N_{u'}={\emptyset}$, then $N_{u}=\emptyset$ and $f(u)=f(u')$.
\end{enumerate}

\end{lemma}

\begin{lemma}
 Let $u$ be a bad word and $|u|=n\geq{L_{\omega}}$, and $u'$ be the ancestor of $u$. Then $u'$ is a bad word and $f(u)=f(u'),m_{u}=m_{u'},n_{u}=n_{u'}$.
\end{lemma}

It is worth noting that from the proof of Lemma 6 it follows that if $u$ is a bad word of length $|u|=n\geq{L_{\omega}}$, then $|u|\equiv{1}(mod\ l)$.

The set of all words of length less than $L_\omega$, having descendants of length at least $L_\omega$ is denoted by $A$.

The set of bad words of length $n$, whose chain of ancestors is $u\rightarrow{u_1}\rightarrow{u_2}\rightarrow{\ldots}\rightarrow{u_{m}=a}$, where $m\in{\mathbb{N}}$ ($m$ is not fixed) and $a\in{A}$, is denoted by $F_{a}^{bad}(n)$. The cardinality of the set $F_{a}^{bad}(n)$ is denoted by $C_{a}^{bad}(n)$.
\begin{lemma}
Let $u\in{F_{a}^{bad}}(n)$, where $n\geq{L_{\omega}}$. Then $f(u)=m_{a}+2n_{a}$.
\end{lemma}

A word $u$ will be called {\em narrow} if its chain of ancestors is $u\rightarrow{\ldots}\rightarrow{u_{k-1}}\rightarrow{u_k}\rightarrow{\ldots}\rightarrow{u_{m}=a}$, where $a\in{A}$, $u_{k}$ is the first bad word in the chain of ancestors, and for the interpretation $\langle{u_{k},i,j}\rangle$ of the word $u_{k-1}$  we have $i+1>l-j$.

\begin{lemma}
Let $u$ be a narrow word with $|u|=n\geq{L_{\omega}}$, $u'$ be an ancestor of $u$, and $u'$ be a bad word. Then $n_{u}=0$ and $f(u)=m_{u'}+n_{u'}$.
\end{lemma}

The set of narrow words of length $n$ whose chain of ancestors is $u\rightarrow{u_1}\rightarrow{u_2}\rightarrow{\ldots}\rightarrow{u_{m}=a}$, where $m\in{\mathbb{N}}$ ($m$ is not fixed), is denoted by $F_{a}^{nar}(n)$. The cardinality of the set $F_{a}^{nar}(n)$ is denoted by $C_{a}^{nar}(n)$.

\begin{lemma}
Let $u\in{F_{a}^{nar}}(n)$, where $|u|=n\geq{L_{\omega}}$. Then $f(u)=m_{a}+n_{a}$.
\end{lemma}

A word $u$ will be called {\em wide} if its chain of ancestors is $u\rightarrow{\ldots}\rightarrow{u_{k-1}}\rightarrow{u_k}\rightarrow{\ldots}\rightarrow{u_{m}=a}$, where $a\in{A}$, $u_{k}$ is the first bad word in the chain of ancestors, and for the interpretation $\langle{u_{k},i,j}\rangle$ of the word $u_{k-1}$ we have $i+1<l-j$.

\begin{lemma}
Let $u$ be a wide word with $|u|=n\geq{L_{\omega}}$, $u'$ be an ancestor of $u$, and $u'$ be a bad word. Then $n_{u}=0$ and $f(u)=m_{u'}+2n_{u'}$.
\end{lemma}

The set of wide words of length $n$ whose chain of ancestors is $u\rightarrow{u_1}\rightarrow{u_2}\rightarrow{\ldots}\rightarrow{u_{m}=a}$, where $m\in{\mathbb{N}}$ ($m$ is not fixed), is denoted by $F_{a}^{wide}(n)$. The cardinality of the set $F_{a}^{wide}(n)$ is denoted by $C_{a}^{wide}(n)$.

\begin{lemma}
Let $u\in{F_{a}^{wide}}(n)$, where $|u|=n\geq{L_{\omega}}$. Then $f(u)=m_{a}+2n_{a}$.
\end{lemma}

\section{Algorithm for finding $f(u)$ }
Suppose that $u$ is a subword of $\omega$ and $|u|=n$.
In this section, we calculate $\sum_{|u|=n}f(u)$.
The set of all subwords of length $n$ of the word $\omega$, whose chain of ancestors is $u\rightarrow{u_1}\rightarrow{u_2}\rightarrow{\ldots}\rightarrow{u_{m}=z}$, where $m\in{\mathbb{N}}$ ($m$ is not fixed), is denoted by $F_{z}(n)$.
The cardinality of the set $F_{z}(n)$ is denoted by $C_{z}(n)$.

We consider the set $A$ introduced in the previous section.
Let $A=A_1\cup{A_2}$ be a partition of set $A$, where $A_1$ is the set of bad words belonging to the set $A$, and $A_2$ is the set of remaining  words of $A$. Thus, for a word $u$ there are two opportunities:

1)$a\in{A_2}$. In this case, Lemma 5  implies that $f(u)=f(a)=m_{a}$,

2)$a\in{A_1}$. In this case, there are two cases:if $u$ is a bad or a wide word, due to Lemmas 7 and 11 we obtain $f(u)=f(a)=m_{a}+2n_{a}$.
If $u$ is a narrow word, then by Lemma 4 we obtain $f(u)=m_{a}+n_{a}$.

\begin{theorem}
$\sum_{|u|=n}f(u)=\sum_{a_1\in{A_1}}[C_{a_1}^{nar}(n)(m_{a_1}+n_{a_1})+(C_{a_1}^{bad}(n)+C_{a_1}^{wide}(n))(m_{a_1}+2n_{a_1})]+\sum_{a_2\in{A_2}}C_{a_2}(n)m_{a_2}$.
\end{theorem}

Now for the calculation of $\sum_{|u|=n}f(u)$ it remains to compute $C_{a}^{nar}(n)$, $C_{a}^{wide}(n)$, $C_{a}^{bad}(n)$, $C_{a}(n)$. Let $n=xl+r$, where $0\leq{r}\leq{l-1}$. It is easy to see that for $xl+r\geq{L_\omega}$ the following recurrence relations hold:

\begin{enumerate}
  \item  $C_{a}^{bad}(xl+1)=l{C_{a}^{bad}(x+1)}$. We note that the remark to Lemma 6 implies that $C_{a}^{bad}(xl+r)=0$ for $r\neq{1}$.
  \item $C_{a}^{nar}(xl+r)=(r-1)C_{a}^{nar}(x+2)+(r-1)C_{a}^{bad}(x+2)+(l-r+1)C_{a}^{nar}(x+1)$ for $r\geq{1}$.

  $C_{a}^{nar}(xl)=(l-1)C_{a}^{nar}(x+1)+(l-1)C_{a}^{bad}(x+1)+C_{a}^{nar}(x)$.
  \item $C_{a}^{wide}(xl+r)=(r-1)C_{a}^{wide}(x+2)+(l-r+1)C_{a}^{wide}(x+1)+\\(l-r+1)C_{a}^{bad}(x+1)$ for $r\geq{2}$.

  $C_{a}^{wide}(xl+1)=l{C_{a}^{wide}(x+1)}$.

  $C_{a}^{wide}(xl)=(l-1)C_{a}^{wide}(x+1)+C_{a}^{wide}(x)+C_{a}^{bad}(x)$.
  \item $C_{a}(xl+r)=(r-1)C_{a}(x+2)+(l-r+1)C_{a}(x+1)$ for $r\geq{1}$.

  $C_{a}(xl)=(l-1)C_{a}(x+1)+C_{a}(x)$.
\end{enumerate}

\section{Special words}

Recall that the subword $v$ of word $\omega$ is called special if $v0$ and $v1$ are also subwords of $\omega$.

Note that the unique interpretation of any special word $v$ of length at least $L_w$ is equal to $\langle{v',i,0}\rangle$.

Indeed, if $j>0$, then $v$ is uniquely complemented to the right to a full block, and thus only one of the words $v0$ and $v1$ is a subword of $\omega$.

\begin{lemma}

Let $(v0,m_1)$ and $(v1,m_2)$ be some occurrences of words $v0$ and $v1$, where $v$ is a special word with the ancestor $v'$.
Let $(v'0,m'_{1})$ and $(v'1,m'_{2})$ be the ancestors of occurrences of $(v0,m_1)$ and $(v1,m_2)$, where $(v'0,m'_{1})$ and $(v'1,m'_{2})$ generate the same permutation.
Then $(v0,m_1)$ and $(v1,m_2)$ also generate the same permutation.
\end{lemma}

\begin{lemma}
Let $(v0,m_1)$ and $(v1,m_2)$ be some occurrences of words $v0$ and $v1$, where $v$ is a special word with the ancestor $v'$.
Let $(v'0,m'_{1})$ and $(v'1,m'_{2})$ be the ancestors of occurrences of $(v0,m_1)$ and $(v1,m_2)$, where $(v'0,m'_{1})$ and $(v'1,m'_{2})$ generate different non-equivalent permutations.
Then $(v0,m_1)$ and $(v1,m_2)$ also generate different non-equivalent permutations.
\end{lemma}

\begin{lemma}
Let $(v0,m_1)$ and $(v1,m_2)$ be some occurrences of words $v0$ and $v1$, where $v$ is a special word with the ancestor $v'$.
Let $(v'0,m'_{1})$ and $(v'1,m'_{2})$ be the ancestors of occurrences of $(v0,m_1)$ and $(v1,m_2)$.
Then the following statements are true:

1)If $(v'0,m'_{1})$ and $(v'1,m'_{2})$ generate equivalent permutations and $|v|=l|v'|$, then $(v0,m_1)$ and $(v1,m_2)$ also generate equivalent permutations.

2)If $(v0,m_1)$ and $(v1,m_2)$ generate equivalent permutations, then $(v'0,m'_{1})$ and $(v'1,m'_{2})$ also generate equivalent permutations and $|v|=l|v'|$.

3)If $(v'0,m'_{1})$ and $(v'1,m'_{2})$ generate equivalent permutations and $|v|<l|v'|$, then $(v0,m_1)$ and $(v1,m_2)$ generate the same permutation.
\end{lemma}

We will consider special word $v$ of length $n-1$.

Let $v$, without loss of generality, start with $0$.
Then $v1$ cannot generate equivalent permutations.

The number of permutations of the set $M_{v0}$ which also belong to $M_{v1}$, is denoted by $k_v$.

The number of permutations of the set $M_{v0}$, such that each of their is equivalent to some permutation of the set $M_{v1}$, is denoted by $t_v$.

The number of pairs of permutations of a set $N_{v0}$ such that a permutation of the pair is equivalent, and the other is equal to some permutation of the set $M_{v1}$, is denoted by $r_v$.

\begin{lemma}

Let $v$ be a special word with the ancestor $v'$, and $|v|<l|v'|$. Then $k_{v}=k_{v'}+t_{v'}+r_{v'}$, $t_{v}=0$ and $r_{v}=0$.

\end{lemma}

\begin{lemma}

Let $v$ be a special word with the ancestor $v'$, and $|v|=l|v'|$. Then $k_{v}=k_{v'}$, $t_{v}=t_{v'}$ and $r_{v}=r_{v'}$.

\end{lemma}

\begin{lemma}

Let $v$ be a special word with the ancestor $v'$, and $t_{v'}=r_{v'}=0$. Then $k_{v}=k_{v'}$ and $t_{v}=r_{v}=0$.

\end{lemma}

\section{Algorithm for finding $g(v)$}
Suppose $v$ is a special word and $|v|=n-1$.
The set of all the special words of length $n$ is denoted by $B(n)$.
The number of common permutations generated by some occurrences of words $v0$ and $v1$ is denoted by $g(v)$.

In this section, we calculate $\sum_{|v|=n-1}g(v)$.

The set of all special subwords of length $n$ of the word $\omega$, whose chain of ancestors is $v\rightarrow{v_1}\rightarrow{v_2}\rightarrow{\ldots}\rightarrow{v_{m}=z}$, is denoted by $B_{z}(n)$.
The cardinality of the set $B_{z}(n)$ is denoted by $S_{z}(n)$.
It is clear that the chain of ancestors of word $v$ consists of special words.
The set of all special words of length less than $L_\omega$, with the descendants of the length greater than $L_\omega$ is denoted by $B$.

\begin{lemma}
Let $xl+r\geq{L_\omega}$, $0\leq{r}\leq{l-1}$, and $b$ be a special word. Then the following recurrence relations hold:

1)$S_{b}(xl+r)=S_{b}(x+1)$ if $r>0$.

2)$S_{b}(xl)=S_{b}(x)$ if $r=0$.

3)$S_{b}(l^{k}|b|)=1$ if $k\geq{1}$.

\end{lemma}

Now we calculate $g(v)$.

\begin{lemma}
Let $v\in{B_{b}(n-1)}$. Then the following statements are true:

1)If $n\neq{l^{k}|b|+1}$  for any positive integer $k$, then $g(v)=k_{b}+t_{b}+r_{b}$.

2)If $n=l^{k}|b|+1$ for some positive integer $k$, then $g(v)=k_{b}+r_{b}$.

\end{lemma}

Let us introduce the function $\delta(n,b)$: if $n=l^{s}|b|+1$ for some positive integer $s$, then $\delta(n,b)=0$, otherwise $\delta(n,b)=1$.

\begin{theorem}
$\sum_{v\in{B(n-1)}}g(v)=\sum_{b\in{B}}[S_{b}(n-1)(k_{b}+t_{b}+r_{b})\delta(n,b)+(k_{b}+r_{b})(1-\delta(n,b))]$.
\end{theorem}

\section{The main theorem}
In this section we state the main theorem of this article.
For this purpose we use the following Lemma (Lemma 1 from $[3]$).

\begin{lemma}
Let $u=u_1\ldots{u_n}$ and $v=v_1\ldots{v_n}$ be two subwords of word $\omega$ and $u_{i}\neq{v_{i}}$ for some $1\leq{i}\leq{n-1}$. Then $u$ and $v$ do not generate the same permutations.

\end{lemma}

We can now prove the main theorem of this paper:

\begin{theorem}
 Let $\omega$ be a fixed point of the morphism $\varphi$, where ${\varphi}\in{Q}$.
Then the permutation complexity of $\omega$ is calculated as follows:
$\lambda(n)=\sum_{a_1\in{A_1}}[C_{a_1}^{nar}(n)(m_{a_1}+n_{a_1})+(C_{a_1}^{bad}(n)+C_{a_1}^{wide}(n))(m_{a_1}+2n_{a_1})]+\sum_{a_2\in{A_2}}C_{a_2}(n)m_{a_2}-\sum_{b\in{B}}[S_{b}(n-1)(k_{b}+t_{b}+r_{b})\delta(n,b)+(k_{b}+r_{b})(1-\delta(n,b))]$.
\end{theorem}

\section{Permutation complexity of the Thue-Morse Word}
In \cite{widmer} Widmer calculated the factor complexity of the permutation generated by the Thue-Morse word.
In this section, we present an alternative proof of the formula for permutation complexity of the Thue-Morse word.
Let $n=2^{k}+b$, where $0<b\leq{2^k}$.
Note that the length $L_\omega$ of the synchronization of the Thue-Morse word is $4$.
It is also easy to understand that $A_1=\{010,101\}$ and $A_2=\{00,01,10,11,001,011,100,101\}$ (for example, $010$ generates two equivalent permutations $132$ and $231$  due to occurrences of words $\omega_{11}\omega_{12}\omega_{13}$ and $\omega_{4}\omega_{5}\omega_{6}$, respectively). Thus, we obtain:

$\sum_{|u|=n}f(u)=\sum_{a_{1}\in{A_1}}[C_{a_1}^{nar}(n)(m_{a_1}+n_{a_1})+(C_{a_1}^{bad}(n)+C_{a_1}^{wide}(n))(m_{a_1}+2n_{a_1})]+\sum_{a_{2}\in{A_2}}C_{a_2}(n)m_{a_2}=
\sum_{a_{1}\in{A_1}}[C_{a_1}^{nar}(n)+2(C_{a_1}^{bad}(n)+C_{a_1}^{wide}(n))]+\sum_{a_{2}\in{A_2}}C_{a_2}(n)=
C_{010}^{bad}(n)+
C_{010}^{wide}(n)+C_{101}^{bad}(n)+C_{101}^{wide}(n)+C(n)$.

It is clear that for $C_{a}^{bad}(n)$ and $C_{a}^{wide}(n)$ the following recurrence relations hold:

$C_{a}^{bad}(2n+1)=2C_{a}^{bad}(n+1)$,
$C_{a}^{wide}(2n+1)=2C_{a}^{wide}(n+1)$, $C_{a}^{wide}(2n)=C_{a}^{wide}(n+1)+C_{a}^{wide}(n)+C_{a}^{bad}(n)$.
Hence it is easy to see that $C_{010}^{bad}(2^{k}+1)=C_{101}^{bad}(2^{k}+1)=2^{k-1}$ for $k>0$; for other $n$ we have
 $C_{010}^{bad}(n)=C_{101}^{bad}(n)=0$.

Let us prove by induction on $n$ that for $2^{k}+2\leq{n}<3*{2}^{k-1}$($k>2$) the relation $C_{010}^{wide}(n)=C_{101}^{wide}(n)=2^{k-1}-b+1$ holds.
The base $n=10$ follows from the relations $C_{010}^{wide}(10)=C_{010}^{wide}(6)+C_{010}^{wide}(5)+C_{010}^{bad}(5)=1+0+2=3=2^{3-1}-2+1$ and  $C_{010}^{wide}(11)=2C_{010}^{bad}(6)=2=2=2^{3-1}-3+1$.

Let us prove the induction step. If $n=2^{k}+2b'$, then by the induction hypothesis we have $C_{010}^{wide}(2^{k-1}+b')=C_{101}^{wide}(2^{k-1}+b')=2^{k-2}-b'+1$ and $C_{010}^{wide}(2^{k-1}+b'+1)=C_{101}^{wide}(2^{k-1}+b'+1)=2^{k-2}-b'$. Hence we obtain $C_{010}^{wide}(n)=C_{101}^{wide}(n)=C_{010}^{wide}(2^{k-1}+b')+C_{010}^{wide}(2^{k-1}+b'+1)+C_{010}^{bad}(2^{k-1}+b')=2^{k}-2b'+1$. If $n=2^{k}+2b'+1$, then by the induction hypothesis we have $C_{010}^{wide}(2^{k-1}+b'+1)=C_{101}^{wide}(2^{k-1}+b'+1)=2^{k-2}-b'$. Hence we obtain $C_{010}^{wide}(n)=C_{101}^{wide}(n)=2C_{010}^{wide}(2^{k-1}+b'+1)=2(2^{k-2}-b')=2^{k-1}-2b'$. The induction step is proved.

Similarly, we prove by induction on $n$ that for $3*2^{k-1}+1\leq{n}\leq{{2}^{k+1}+1}$ the relation $C_{010}^{wide}(n)=C_{101}^{wide}(n)=0$ holds. The base $n=7$ and $n=8$ follow from the relations $C_{010}^{wide}(8)=C_{101}^{s}(8)=C_{010}^{wide}(5)+C_{010}^{wide}(4)+C_{010}^{bad}(4)=0+0+0=0$, $C_{010}^{wide}(7)=C_{101}^{wide}(7)=2C_{010}^{wide}(4)=0$. Let us prove the induction step. If $n=2^{k}+2b'$, then by the induction hypothesis we have $C_{010}^{wide}(2^{k-1}+b')=C_{101}^{wide}(2^{k-1}+b')=0$ and $C_{010}^{wide}(2^{k-1}+b'+1)=C_{101}^{s}(2^{k-1}+b'+1)=0$. Hence we obtain $C_{010}^{wide}(n)=C_{101}^{wide}(n)=C_{010}^{wide}(2^{k-1}+b')+C_{010}^{wide}(2^{k-1}+b'+1)+C_{010}^{bad}(2^{k-1}+b'+1)=0+0+0=0$. If $n=2^{k}+2b'+1$, then by the induction hypothesis we have $C_{010}^{wide}(2^{k-1}+b'+1)=C_{101}^{wide}(2^{k-1}+b'+1)=0$.  Hence we obtain $C_{010}^{wide}(n)=C_{101}^{wide}(n)=2C_{010}^{wide}(2^{k-1}+b'+1)=2\cdot{0}=0$. The induction step is proved.

Thus we have proved the following:

1) If $2^{k}+2\leq{n}<3*{2}^{k-1}$, then $\sum_{|u|=n}f(u)=2(2^{k-1}-b+1)+4(2^{k}+b-1)-2^{k}=2^{k+2}+2b-2$.

2) If $3*2^{k-1}+1\leq{n}<{2}^{k+1}+1$, then $\sum_{|u|=n}f(u)=2(n-1)+2^{k+1}=2^{k+2}+2b-2$.

Therefore the equality $\sum_{|u|=n}f(u)=C(n)=2(n-1)+2^{k+1}=2^{k+2}+2b-2$ holds for all $n\geq{6}$.

It is clear that for $S_{b}(n)$ the following recurrence relations hold:

$S_{b}(2n+1)=S_{b}(n+1)$ and $S_{b}(2n)=S_{b}(n)$.

Taking into account that $S_{01}(3)=S_{10}(3)=1$ and $S_{01}(4)=S_{10}(4)=1$, it is easy to prove by induction that for $n\geq{3}$  the equality $S_{01}(n)=S_{10}(n)=1$ holds.

It is also easy to see that $B=\{01,10,010,101\}$.

In addition word $010$ generates two equivalent permutations $132$ and $231$, while word $011$ generates only a single permutation $132$.
Hence $k_{01}=t_{01}=0$ and $r_{01}=1$. Similarly $k_{10}=t_{10}=0$ and $r_{10}=1$.

Words $0101$ and $0100$ generate different inequivalent permutations $1324$ and $3421$.
Hence $k_{010}=t_{010}=0$ and $r_{010}=0$. Similarly $k_{101}=t_{101}=0$ and $r_{101}=0$. Therefore for $n\neq{2^k}+1$ by Theorem 2 the following relation holds:

$\sum_{v\in{B(n-1)}}g(v)=\sum_{b\in{B}}[S_{b}(n-1)(k_{b}+t_{b}+r_{b})\delta(n,b)+(k_{b}+r_{b})(1-\delta(n,b))]=
S_{01}(n-1)(k_{01}+t_{01}+r_{01})+S_{10}(n-1)(k_{10}+t_{10}+r_{10})=1+1=2$.

For $n=2^{k}+1$ by Theorem 2 the following relation holds:

$\sum_{v\in{B(n-1)}}g(v)=\sum_{b\in{B}}[S_{b}(n-1)(k_{b}+t_{b}+r_{b})\delta(n,b)+(k_{b}+r_{b})(1-\delta(n,b))]=
k_{01}+r_{01}+k_{10}+r_{10}=2$.

Thus the formula for the permutation complexity of the Thue-Morse word is

$\lambda(n)=\sum_{|u|=n}f(u)-\sum_{b\in{B(n-1)}}g(b)=2^{k+2}+2b-2-2=2(2^{k+1}+b-2)$ for $n=2^{k}+b$, where $0<b\leq{2^k}$.

\section{Acknowledgements}
I am grateful to A. E. Frid and S. V. Avgustinovich for helpful and stimulating discussions.

\bibliographystyle{eptcs}
\bibliography{val}

\begin{thebibliography}{1}
\providecommand{\bibitemdeclare}[2]{}
\providecommand{\urlprefix}{Available at }
\providecommand{\url}[1]{\texttt{#1}}
\providecommand{\href}[2]{\texttt{#2}}
\providecommand{\urlalt}[2]{\href{#1}{#2}}
\providecommand{\doi}[1]{doi:\urlalt{http://dx.doi.org/#1}{#1}}
\providecommand{\bibinfo}[2]{#2}

\bibitemdeclare{article}{flaas}
\bibitem{flaas}
\bibinfo{author}{D.~G. Fon-Der-Flaass} \& \bibinfo{author}{A.~E. Frid}
  (\bibinfo{year}{2007}): \emph{\bibinfo{title}{On periodicity and low
  complexity of infinite permutations}}.
\newblock {\sl \bibinfo{journal}{European J. Combin.}}
  \bibinfo{volume}{28}(\bibinfo{number}{8}), pp. \bibinfo{pages}{2106--2114},
  \doi{10.1016/j.ejc.2007.04.017}.

\bibitemdeclare{phdthesis}{frid}
\bibitem{frid}
\bibinfo{author}{A.E. Frid} (\bibinfo{year}{2000}): \emph{\bibinfo{title}{On
  Combinatorial Properties of fixed points of morphisms}}.
\newblock Ph.D. thesis, \bibinfo{school}{Ph.D. thesis, Novosibirsk}.

\bibitemdeclare{article}{mak}
\bibitem{mak}
\bibinfo{author}{M.~A. Makarov} (\bibinfo{year}{2006}):
  \emph{\bibinfo{title}{On permutations generated by infinite binary words}}.
\newblock {\sl \bibinfo{journal}{Sib. \`Elektron. Mat. Izv.}}
  \bibinfo{volume}{3}, pp. \bibinfo{pages}{304--311 (electronic)}.

\bibitemdeclare{article}{sturm}
\bibitem{sturm}
\bibinfo{author}{M.~A. Makarov} (\bibinfo{year}{2009}):
  \emph{\bibinfo{title}{On permutations generated by {S}turmian words}}.
\newblock {\sl \bibinfo{journal}{Sibirsk. Mat. Zh.}}
  \bibinfo{volume}{50}(\bibinfo{number}{4}), pp. \bibinfo{pages}{850--857},
  \doi{10.1007/s11202-009-0076-6}.

\bibitemdeclare{article}{widmer}
\bibitem{widmer}
\bibinfo{author}{S~Widmer} (\bibinfo{year}{2011}):
  \emph{\bibinfo{title}{Permutation complexity of the Thue-Morse word}}.
\newblock {\sl \bibinfo{journal}{Advances in Applied Mathematics}}
  \bibinfo{volume}{47}(\bibinfo{number}{2}), pp. \bibinfo{pages}{309--329},
  \doi{10.1016/j.aam.2010.08.002}.

\end{thebibliography}

%\begin{thebibliography}{6}
%\bibitem{avg}
%S. V . Avgustinovich. The number of distinct subwords of fixed length in the Morse-Hedlund sequence. Sibirsk. zhurnal issledovaniya operatsii.
% 1 no. 2(1994) 3-7
% 
%\bibitem{frid}
%A. E.  Frid. On Combinatorial Properties of fixed points of morphisms. Ph.D thesis. Novosibirsk, 2000.
%
%\bibitem{mak}
%M.A. Makarov. On permutations generated by infinite binary words. Sib. Elektron. Mat. Izv., 3:304-311,
%2006. (in Russian).
%
%\bibitem{flaas}
%D.G. Fon-Der-Flaass and A.E. Frid. On periodicity and low complexity of infinite permutations. European
%J. Combin., 28(8):2106-2114, 2007.
%
%\bibitem{sturm}
%M.A. Makarov. On the permutations generated by the Sturmian words. Sib. Math. J., 50(3):674-680,
%2009.
%
%\bibitem{widmer}
%S. Widmer. Permutation complexity of the Thue-Morse word. Adv. in Appl. Math.
%2010.
%
%\end{thebibliography}

\end{document}